\documentclass[twocolumn,twocolappendix]{aastex701}
\usepackage{xspace}

\usepackage{CJKutf8}
\usepackage{amsmath}
\usepackage{hyperref}
\usepackage{tikz}
\usetikzlibrary{arrows.meta,decorations.markings,patterns,calc,positioning,fit,matrix}

\begin{document}
\begin{CJK*}{UTF8}{gbsn}

\shorttitle{Exomoons Around Migrating White Dwarf Planets}
\shortauthors{Narayan, Livesey \& Becker}

\title{Exomoonlight Sonata: \\ On the Fate of Exomoons Around Migrating White Dwarf Planets}

\author[orcid=0009-0007-0488-5685]{Ritvik Sai Narayan}
\altaffiliation{Thomas Frank \& Henry W. Kendall Fellow}
\affiliation{Department of Physics, Massachusetts Institute of Technology, Cambridge, MA 02139, USA}
\affiliation{MIT Kavli Institute for Astrophysics and Space Research, Massachusetts Institute of Technology, Cambridge, MA 02139, USA}
\affiliation{Department of Astronomy, University of Wisconsin--Madison, 475 N.~Charter St., Madison, WI 53706, USA}
\affiliation{Wisconsin Center for Origins Research, University of Wisconsin--Madison, 475 N Charter St, Madison, WI 53706, USA}
\email[show]{ritviksn@mit.edu}  

\author[0000-0003-3888-3753]{Joseph R. Livesey}
\affiliation{Department of Astronomy, University of Wisconsin--Madison, 475 N.~Charter St., Madison, WI 53706, USA}
\affiliation{Wisconsin Center for Origins Research, University of Wisconsin--Madison, 475 N Charter St, Madison, WI 53706, USA}
\email{jrlivesey@wisc.edu}

\author[0000-0002-7733-4522]{Juliette Becker}
\affiliation{Department of Astronomy, University of Wisconsin--Madison, 475 N.~Charter St., Madison, WI 53706, USA}
\affiliation{Wisconsin Center for Origins Research, University of Wisconsin--Madison, 475 N Charter St, Madison, WI 53706, USA}
\email{juliette.becker@wisc.edu}

\begin{abstract}
The survival of exomoons around close-in white dwarf planets offers a potential constraint on their migration histories and an alternative pathway to post-main-sequence habitability. We investigate whether moons can survive the tidal circularization of planets via high-eccentricity migration, and consider a hierarchical three-body system (white dwarf--planet--moon) after an initial eccentricity excitation of the planetary orbit. Using an analytic equilibrium tidal model, we compare the evolution of the planetary and satellite orbits and examine the conditions required for an evection resonance. We show that the moons of pseudosynchronously rotating planets always migrate inwards since the planetary corotation radius always exceeds the Hill radius, independent of mass and orbital eccentricities. Comparing the migration timescales for the moon and the planet shows that compact moons undergo rapid tidal decay relative to the outer orbit. However, the wider separations favored by slower migration timescales experience pericenter perturbations whose work constitutes a substantial fraction of the moon's orbital energy. Together, our results indicate that moons likely do not survive during a planet's high-eccentricity migration phase in a three-body configuration. 
\end{abstract}


\section{Introduction} \label{section:introduction}

The discovery that planetary systems can survive after their host stars leave the main sequence has reframed white dwarfs (WDs) as fascinating laboratories for the final stages of planetary evolution. At least a quarter, and possibly up to half, of WDs carry photospheric metals that are thought to be continuously resupplied by infalling planetary material \citep{koester_frequency_2014}; a subset host dusty and gaseous debris disks \citep{farihi_infrared_2009}; and one system WD 1145+017 shows transiting fragments of disintegrating planetesimals near the tidal-disruption radius \citep{vanderburg_disintegrating_2015}. Planetary collisions following asymptotic giant branch mass loss may also generate rocky debris that subsequently migrates inward under Poynting--Robertson drag \citep{adams_production_2026}, providing a possible additional source of WD metal pollution. Although limited in sample size, intact planets have been found as well. Prominent examples include a possible evaporating ice giant orbiting WD J0914+1914 \citep{gansicke_accretion_2019,brouwers_asynchronous_2023}, the transiting giant planet WD 1856+534\,b \citep{vanderburg_giant_2020}, whose planetary nature was confirmed by JWST/MIRI thermal emission at a temperature of $186^{+6}_{-7}$ K, constraining its mass to $\lessapprox 6 \, {\rm M_{J}}$ \citep{limbach_thermal_2025}, and a $1.4\, {\rm M_J}$ companion on a wide orbit  ($\gtrsim 2.8$ AU) around a $0.53 \, {\rm M_\odot}$ WD detected via microlensing \citep{blackman_jovian_2021}. Collectively, these detections establish that giant planets can survive stellar mass loss, and in some cases, arrive on close orbits around the WD. 

The inward migration of giant planets to stable orbits opens the possibility of their moons arriving and residing within the WD habitable zone as well. Since WDs cool slowly over billions of years, bodies that migrate close enough could experience temperate surface conditions long enough to support habitable conditions \citep{fossati_habitability_2012,barnes_habitable_2013}. Combined transit timing and duration variations offer a route to detecting and characterizing exomoons around main-sequence stars \citep{kipping_transit_2009}, with early forecasts suggesting that sufficiently massive habitable-zone moons could be detected with Kepler-quality photometry \citep{kipping_detectability_2009}. Subsequent searches have identified candidates such as Kepler-1708\,b-i, underscoring both the promise of these observations and the need for further measurements to establish their moon-like nature \citep{kipping_exomoon_2022}. 

In this effort, next-generation observatories will be essential for probing post-main-sequence habitability scenarios, particularly in determining whether exoplanets that survive their host star's red giant phase can develop habitable conditions \citep{payne_liberating_2016,becker_fate_2025}. Crucially, the temperate exosatellites that accompany such planets may soon be within observational reach with future flagship missions. Recent methodologies such as the lunar eclipse technique detailed by \citet{limbach_exomoons_2026} demonstrate that the \textit{Habitable Worlds Observatory} \citep{clery_future_2023} could identify Earth-sized exomoons orbiting giant planets via broadband reflected-light lunar eclipses, providing a direct mechanism to investigate this pathway to habitability. 

However, retaining a moon during planetary migration to the WD habitable zone is a dynamically challenging process \citep{spalding_resonant_2016}. Giant planets that survive post-main-sequence phases do so at large orbital distances, where adiabatic mass loss expands their orbital periods by a factor of few \citep{villaver_can_2007,mustill_foretellings_2012}. Delivering such a planet to a close orbit around the WD thus requires a late phase of high-eccentricity migration where the planet is driven to a highly eccentric orbit via mechanisms such as planet--planet scattering \citep[e.g.,][]{rasio_dynamical_1996,weidenschilling_gravitational_1996,lin_origin_1997,chatterjee_dynamical_2008}, secular forcing from a companion \citep[e.g.,][]{laskar_large_1997,laskar_chaotic_2008,lithwick_theory_2011,lithwick_secular_2014,oconnor_secular_2022}, or a WD kick driven by asymmetric mass loss \citep[e.g.,][]{spruit_origin_1998,heyl_orbital_2007,batygin_terminal_2026}. Following this excitation, the orbit is subsequently circularized by tidal dissipation during repeated pericenter passages \citep{veras_tidal_2019,munoz_kozai_2020,oconnor_enhanced_2021}. 

Recent simulations of the analogous main-sequence problem considering giant planet migration via von Zeipel-Lidov-Kozai \citep[vZLK;][]{von_zeipel_sur_1910,lidov_evolution_1962,kozai_secular_1962} oscillations find that moons are almost universally lost \citep{trani_ominous_2020}. A moon that does remain bound faces the threat of its own tidal interaction with its host planet, which over Gyr timescales could carry a moon to the Roche limit or beyond the Hill radius \citep{barnes_stability_2002,alvarado-montes_effect_2017,sucerquia_can_2020,tokadjian_impact_2020}. 

We investigate this second phase of tidal evolution in the context of three-body WD systems. In this effort, we examine the phase of dynamical evolution post-eccentricity excitation to determine if a moon that survived the previous phases could survive this final phase of evolution. 

We examine three processes that constrain satellite survival: (1) \textit{tidal migration}, which can drive compact moons inward faster than the planet migrates, (2) \textit{evection resonance}, which can excite satellite eccentricity in certain configurations; and (3) \textit{gravitational perturbations} during WD pericenter passages, which most strongly affect wider moons. These effects impose competing constraints: increasing a moon's separation slows its tidal evolution but strengthens its exposure to stellar perturbations, while planetary migration can lead to a resonance. For a moon to survive the tidal migration of its host giant planet, it must therefore reside in a region of parameter space where the tidal decay of its orbit is sufficiently slow, resonantly driven growth of orbital eccentricity is avoided, and kicks at periastron do not destabilize its orbit. In this Letter, we evaluate the interplay among these three mechanisms to determine whether a viable region of parameter space exists in which such moons can survive.

This Letter is organized as follows. Section~\ref{sec:methods} presents our analytic tidal model and evection resonance criterion. Section~\ref{sec:results} examines satellite survival through resonance encounters, tidal migration, and perturbations during WD pericenter passages. In Section~\ref{sec:discussion}, we compare our results with previous work and discuss their implications for planetary migration histories and the prospects for habitable moons around WDs. 

\section{Methods} \label{sec:methods}

\subsection{Analytic Tidal Model} \label{subsec:tidal_model}

We consider a hierarchical system consisting of a WD of mass $M_\star$, a planet of mass $M_{\rm p}$, and a moon of mass $M_{\rm m}$. The WD-planet and planet-moon orbits are described by the elements ($a_{\rm p}$, $e_{\rm p}$) and ($a_{\rm m}$, $e_{\rm m}$), respectively, corresponding to their semi-major axes and eccentricities. We initially take the moon's orbit to be circular ($e_{\rm m} \simeq 0$) and the mutual inclination between orbits and spins to be either aligned (prograde) or anti-aligned (retrograde) such that $\varepsilon_{\rm p} \approx 0$ and $\varepsilon_{\rm m} \approx 0, \pi$. This choice is motivated by the need to prevent the dynamical instabilities driven due to vZLK oscillations in circular satellite orbits with $\varepsilon_{\rm m} \gtrapprox 39^\circ$ as demonstrated by \citet{lidov_evolution_1962} (see Section 7, Case 2 by \citeauthor{lidov_evolution_1962} for the ``vertical moon hypothesis''; \citealt{tremaine_why_2014}). 

We adopt $M_\star = 0.5 \, M_\odot$, as WDs descended from Sun-like stars typically retain only half their progenitor mass following the post-main-sequence phases \citep{schroder_distant_2008,cummings_white_2018,batygin_terminal_2026}. Our fiducial companion has $M_{\rm p} = 10 \; M_{\rm J}$ and $R_{\rm p} = 1.2\; R_{\rm J}$, while the moon has $M_{\rm m} = 0.025 \; M_\oplus$ and $R_{\rm m} = 0.413 \; R_\oplus$, typical of a Ganymede-like moon \citep{heller_exomoon_2013}. The outer orbit initially begins at $a_{\rm p,0} = 4.5$\;AU and $e_{\rm p, 0} = 0.995$, implying a circularized separation of ${\simeq}0.045$\;AU under approximate conservation of angular momentum. Although this lies at the outer edge of the continuously habitable zone (at early times) identified by \citet{agol_transit_2011}, the wider initial pericenter provides a favorable configuration for moon retention by allowing a larger Hill radius. We subsequently vary the lunar orbital separation, parameterizing it by the planet's Hill radius at pericenter, 
\begin{equation} \label{eqn:a_m}
    a_{\rm m} = f \cdot R_{\rm Hill, \, peri},
\end{equation}
where
\begin{equation}
   R_{\rm Hill, \, peri} \simeq a_{\rm p} (1-e_{\rm p}) \left( \frac{M_{\rm p}}{3M_\star} \right)^{1/3},
\end{equation}
and $0< f< 1$. 

\subsubsection{Tidal Lag Prescription}

In the weak-friction approximation of equilibrium tidal theory \citep{zahn_tidal_2008}, the tidal deformation can be assumed to follow the companion's time-varying potential with a constant time lag (CTL) model, with $\tau_i^j$ as the lag parameter where the subscript $i$ denotes the deformed body and the superscript $j$ denotes the tidal companion. Since the planet has two tidal partners, tides raised on the planet have contributions from both the WD and the moon. Given that the CTL model assigns only a single value of $\tau_{\rm p}$ in its default parameterization, we specify this lag through a reference tidal quality factor, $Q_{\rm p}$, that varies for the inner and the outer orbits. 

Eccentric tidal forcing often contains a spectrum of harmonics, whose phase lags and amplitudes are sensitive to the forcing frequency \citep{efroimsky_tidal_2013,efroimsky_tidal_2014}. We adopt the standard weak-friction approximation, in which the harmonics of each tidal interaction share a CTL, retaining the orbit-averaged equations \citep{hut_tidal_1981,leconte_is_2010}. Following common prescriptions in tidal evolution calculations, we parameterize the time lag through a reference tidal quality factor and a characteristic frequency \citep[e.g.,][]{leconte_is_2010,lu_self-consistent_2023,millholland_spin_2024}. For the highly eccentric WD--planet orbit, we adopt a pericenter-based reference frequency, motivated by the concentration of tidal forcing near pericenter. Modifying Equation (2.32) of \citet{murray_solar_1999}, we have that the Keplerian relation for the outer orbit is, 
\begin{equation}
    r_{\rm p} \dot{\nu}_{\rm p} = \frac{n_{\rm p}a_{\rm p}}{\sqrt{1-e^2_{\rm p}}} \left(1+e_{\rm p}\cos \nu_{\rm p} \right)
\end{equation}
where $\nu_{\rm p}$ is the true anomaly of the planet and 
\begin{equation}
    n_{\rm p} = \sqrt{\frac{G(M_\star + M_{\rm p})}{a_{\rm p}^3}}
\end{equation}
is the mean angular frequency of the orbit. At periastron, $\nu_{\rm p} = 0$ and $r_{\rm p} = a_{\rm p} (1-e_{\rm p})$, yielding, 
\begin{equation}
    \dot{\nu}_{\rm p, \, peri} = n_{\rm p} \frac{(1+e_{\rm p})^{1/2}}{(1-e_{\rm p})^{3/2}} \equiv \Gamma(e_{\rm p})  \cdot n_{\rm p}.
\end{equation}

To define a single time lag over the migration interval, we adopt the weighted time average of the dimensionless factor $\Gamma$, 
\begin{equation} \label{eqn:Gamma}
\begin{split}
\left\langle \Gamma \right\rangle
& =
\frac{
\displaystyle \int_{t_i}^{t_f}
\frac{[1+e_{\rm p}(t)]^{1/2}}{[1-e_{\rm p}(t)]^{3/2}}\;dt
}{
t_f-t_i
} \\
& =
\frac{
\displaystyle \int_{e_f}^{e_i}
\frac{(1+e_{\rm p})^{1/2}}{(1-e_{\rm p})^{3/2}}
\frac{1}{|\dot{e}_{\rm p}|}\;de_{\rm p}}
{\displaystyle \int_{e_f}^{e_i} \frac{1}{|\dot{e}_{\rm p}|}\;de_{\rm p}}.
\end{split}
\end{equation}
The $1/|\dot{e}_{\rm p}|$ weights each eccentricity interval by the time spent there. Then, using the standard eccentricity polynomials, $f_1(e^2)$--$f_5(e^2)$, from \citet{hut_tidal_1981}, Equation~\eqref{eqn:Gamma} can be recast as,
\begin{equation} 
    \langle \Gamma \rangle = 
    \frac{
     \displaystyle \int_{0.01}^{0.995} \frac{f_5}{e_{\rm p}(1-e_{\rm p})^3 (1+e_{\rm p}) [18f_3 f_5 - 11 f_2 f_4]} \, de
    }{ 
     \displaystyle \int_{0.01}^{0.995}\frac{f_5 }{e_{\rm p}(1-e_{\rm p}^2)^{3/2} [18f_3 f_5 - 11 f_2 f_4]} \, de
    }
\end{equation}
since all constant factors in the eccentricity-damping rate cancel out between the numerator and the denominator. We augment this definition with the commonly used prescription of $\tau \sim (2nQ)^{-1}$ \citep[see][]{lu_self-consistent_2023} describing a semi-diurnal tide in the limit of a circular, synchronous system. This yields the planetary lag for the outer orbit as, 
\begin{equation}
    \tau_{\rm p}^{\rm WD}
    \sim \frac{1}
    {2Q_{\rm p} \langle \Gamma \rangle n_{\rm p, \, 0}},
    \label{eq:planet_time_lag}
\end{equation}

For a circular, coplanar prograde moon--planet orbit, we evaluate the effective tidal lag raised on the companion at the semidiurnal forcing frequency, 

\begin{equation}
    \tau_{\rm p}^{\rm m} \sim \frac{1}{2Q_{\rm p}|n_{\rm m, \, 0} - \Omega_{\rm p, \, 0}|}, 
\end{equation}
and
\begin{equation}
        \tau_{\rm m}^{\rm p} \sim \frac{1}{2Q_{\rm m} n_{\rm m, 0}}.
\end{equation}
We adopt reference tidal quality factors $Q_{\rm p} =10^6$ and $Q_{\rm m}= 10^2$ to normalize our tidal lag prescriptions. These values follow typical choices used in previous exomoon studies: \citet{tokadjian_impact_2020} adopt $Q = 10^6$ for gas giants, while \citet{forgan_exomoon_2016} used $Q =100$ in their modeling of terrestrial satellites. 

\subsubsection{Tidal Evolution Equations}

We describe the tidal evolution of this system using the equilibrium-tide model of \citet{hut_tidal_1981}, as reexpressed by \citet{leconte_is_2010}. For a two-body system with masses $m_i$ and $m_j$, semi-major axis $a$, eccentricity $e$, and mean motion $n$, the orbit-averaged tidal equations in the limit of aligned spins are given by,
\begin{align} 
    \dot{a} & = \frac{4a^2}{Gm_im_j} \sum_i K_{i,j} \left( N(e) \frac{\Omega_i}{n} - N_a(e)\right), \label{eqn:tidal_migration_a} \\ 
    \dot{e} & = \frac{11 ae}{Gm_im_j} \sum_i K_{i,j} \left( \omega_e(e) \frac{\Omega_i}{n} - \frac{18}{11} N_e(e) \right), \label{eqn:tidal_migration_e} \\
    \dot{\Omega}_i & = -\frac{2K_{i,j}}{I_in} \left( \omega(e) \frac{\Omega_i}{n} - N(e) \right), \label{eqn:tidal_migration_omega}
\end{align}
where $\Omega_i$ is the spin of the deformed body and $I_i$ is its moment of inertia (which is assumed to be constant in the derivation of these equations). The coefficient describing dissipation within a body $i$ is given by, 
\begin{equation}
    K_{i,j} = \frac{3}{2} k_{2, \, i} \tau_i \left( \frac{Gm_i^2}{R_i} \right) \left( \frac{m_j}{m_i} \right)^2 \left( \frac{R_i}{a} \right)^6 n^2, \qquad i \neq j,
\end{equation}
where $R_i$ and $k_{2, \, i}$ are the body's radius and degree-two Love number, respectively. We adopt constant values of $k_{2,\, \rm p}=k_{2,\, \rm m}=0.3$. We refer the reader to \citet{leconte_is_2010} for a more detailed description of the eccentricity functions, $N(e)$, $N_a(e)$, $\omega_e (e)$, $N_e(e)$, and $\omega(e)$. 

Given that the equilibrium tides framework describes the evolution of a two-body system, we apply Equations~\eqref{eqn:tidal_migration_a}--\eqref{eqn:tidal_migration_omega} separately to the WD--planet and planet--moon orbits, using the aforementioned tidal lags. We include tides raised on the planet by the WD and moon, and tides raised on the moon by the planet. We neglect dissipation within the WD given that planetary tides dominate in this system \citep{campbell_tidal_1984,willems_energy_2010,millholland_empirical_2025}. 

\subsubsection{Rotation Frequencies}

We take the planet's rotation to be pseudosynchronous with the WD--planet orbit, which is defined as, 
\begin{equation} 
\begin{split}
    \Omega_{\rm ps} & = n_{\rm p} \frac{N(e_{\rm p})}{\omega(e_{\rm p})} \\
    & = n_{\rm p} \frac{1 + \frac{15}{2} e_{\rm p}^2 + \frac{45}{8}e_{\rm p}^4 + \frac{5}{16} e_{\rm p}^6}{(1 + 3e_{\rm p}^2 + \frac{3}{8} e_{\rm p}^4)(1-e_{\rm p}^2)^{3/2}}  
\end{split}
\end{equation}
and we have that $\Omega_{\rm ps} \to n_p$ as the orbit circularizes. This choice is motivated by the tendency of the tidal torque to drive the planetary spin toward a stable equilibrium. For the tides raised on the planet by the WD, the spin-evolution equation can be recast as, 
\begin{equation}
    \dot{\Omega}_{\rm p} = - \frac{\Omega_{\rm p} - \Omega_{\rm ps}}{t_{\rm spin}},
\end{equation}
where 
\begin{equation}
    t_{\rm spin} = \frac{I_{\rm p}n_{\rm p}^{2}}
{2K_{{\rm p},\star}\omega(e_{\rm p})}. 
\end{equation}
For fixed orbital elements, departures from $\Omega_{\rm ps}$ therefore decay exponentially on the spin timescale, $t_{\rm spin}$. Since orbital evolution occurs throughout this relaxation, we have that $\Omega_{\rm p} < \Omega_{\rm ps}$ enhances inward migration and eccentricity damping relative to their pseudosynchronous rates. Conversely, a faster spin can reduce these rates, and if sufficiently rapid can initially drive orbital expansion and eccentricity growth \citep[cf. Figure 1 of][]{leconte_is_2010}. The corresponding thresholds are,
\begin{subequations}
    \begin{equation}
        \Omega_{{\rm p, \, }\dot{a} = 0} = n_{\rm p}\frac{N_a(e_{\rm p})}{N(e_{\rm p})},
    \end{equation}
    \begin{equation}
        \Omega_{{\rm p, \, }\dot e=0}
=\frac{18}{11}n_{\rm p}
\frac{N_e(e_{\rm p})}{\omega_e(e_{\rm p})}.
    \end{equation}
\end{subequations}

\noindent For an eccentric orbit, both thresholds exceed $\Omega_{\rm ps}$. We therefore adopt pseudosynchronous rotation since it avoids selecting an initial spin that temporarily enhances or suppresses orbital decay. Similarly, we also take the moon to rotate synchronously with its initial circular orbit, $\Omega_{\rm m} = n_{\rm m}$, which is the zero-obliquity spin equilibrium for a circular orbit. Since dissipation within the WD is neglected, its rotation does not enter our tidal model. Lastly, we write each body's moment of inertia as $I_i = \alpha_i M_i R_i^2$, where the dimensionless coefficient $\alpha_i$\footnote{The tidal formalism by \citet{hut_tidal_1981} uses the radius of gyration, $r_g$, which is $\sqrt{\alpha}$.} describes its internal mass distribution. We adopt $\alpha_{\rm p} = 0.25$ and $\alpha_{\rm m} =0.33$, approximately corresponding to the Jupiter and Earth values, respectively \citep{williams_jupiter_2016}. 

\subsection{Evection Resonance Criterion} \label{subsec:evection_criterion}

Evection resonance provides a mechanism through which stellar perturbations can excite a satellite's orbital eccentricity to large values, with this mechanism being extensively investigated in the early evolution of the Earth--Moon system \citep[e.g.,][]{touma_resonances_1998,rufu_did_2019}. The resonance occurs when the satellite's apsidal precession rate becomes commensurate with the planet's mean motion around the star. For a coplanar orbit whose precession rate is dominated by planetary oblateness, the apsidal precession frequency is approximately \citep{danby_fundamentals_1970, spalding_resonant_2016},
\begin{equation}
    \dot{\varpi}_{\rm m} \approx \frac{3}{2} J_2 \left( \frac{R_{\rm p}}{a_{\rm m}} \right)^2 n_{\rm m} \frac{1}{(1-e_{\rm m}^2)^2} 
\end{equation}
where $\varpi_{\rm m}$ is the satellite's longitude of pericenter and $J_2$ the is quadrupole moment of the planet. 

For migrating planets, evection offers a potential pathway to satellite loss. \citet{spalding_resonant_2016} demonstrated that inward planetary migration can lead to resonant capture and continued growth of a moon's eccentricity, ultimately leading to a collision with the planet. As the satellite eccentricity increases, its pericenter distance decreases at a given semi-major axis. Sufficient excitation can therefore bring the moon within the planet's Roche limit, eventually leading to destruction. 

Motivated by this mechanism, we investigate where the evolving WD--planet--moon system could encounter an evection resonance during tidal migration.  While we identify candidate configurations for resonant excitation, whether substantial eccentricity growth occurs can also depend on factors such as tidal damping and the possibility of capture \citep{touma_resonances_1998,xu_disruption_2016}. 

Supposing that the moon has a low eccentricity orbit ($e_{\rm m} \approx 0$), we see that the resonance condition is,
\begin{equation}
    \frac{3}{2} J_2 \left( \frac{R_{\rm p}}{a_{\rm m}} \right)^2 n_{\rm m} \approx \left( \frac{GM_\star}{a_{\rm p}^3} \right)^{1/2}.
\end{equation}
Rearranging this expression yields the resonance crossing relation in Equation (5) of \citet{spalding_resonant_2016}, 
\begin{equation}
    a_{\rm res} = R_{\rm p} \left[ \left( \frac{2}{3J_2} \right)^2 \left( \frac{a_{\rm m}}{R_{\rm p}} \right)^7 \left( \frac{M_\star}{M_{\rm p}} \right) \right]^{1/3},
\end{equation}
with $a_{\rm res}$ corresponding to the (astrocentric) semi-major axis of the planet that is required for a resonance. Using the parameterized semi-major axis of the moon defined in Equation~\eqref{eqn:a_m},
\begin{equation}
    a_{\rm res} = \left(\frac{3^{3/2}J_2}{2} \right)^{1/2} \left( \frac{3M_\star}{M_{\rm p}} \right)^{1/3} \frac{R_{\rm p}}{\left[ f(1-e_{\rm p})\right]^{7/4}}.
\end{equation}
From Equations (2) and (3) of \citet{ragozzine_probing_2009}, we can see that the oblateness factor, $J_2$, can be split into having contributions from rotation and tides, 
\begin{equation}
    \begin{split}
        J_2 & = J_{\rm 2, \, spin} + \langle J_{\rm 2, \, tidal} \rangle \\
        & =  \frac{k_2}{3} \left[ \frac{\Omega_{\rm p}^2 R^3_{\rm p}}{GM_{\rm p}} + \frac{3}{2} \left(\frac{R_{\rm p}}{a_{\rm res} } \right)^3 \left( \frac{M_\star}{M_{\rm p}} \right) \frac{1}{(1-e_{\rm p}^2)^{3/2}}\right].
    \end{split}
\end{equation}
where we have orbit-averaged $J_{2, \rm tidal}$ since it depends on the instantaneous orbital separation. Solving for $a_{\rm res}$ with the new $J_2$ yields a quintic equation, 
\begin{equation} \label{eqn:quintic}
    a^5_{\rm res} - \Lambda_1  a_{\rm res}^3 - \Lambda_2 = 0
\end{equation}
with $\Lambda_1, \, \Lambda_2 \in \mathbb{R}^+$ defined as, 

\begin{subequations}
\begin{align}
    \Lambda_1 & = \frac{3\cdot 3^{1/6}\,k_2 M_\star^{2/3}\Omega_{\rm p}^{2}R_{\rm p}^{5}}
{2GM_{\rm p}^{5/3}[f(1-e_{\rm p})]^{7/2}}, \\
 \Lambda_2  & = \frac{9\cdot 3^{1/6}\,k_2 M_\star^{5/3}R_{\rm p}^{5}}
{4M_{\rm p}^{5/3}[f(1-e_{\rm p})]^{7/2}(1-e_{\rm p}^2)^{3/2}}. 
\end{align}
\end{subequations}

\noindent Combining Descartes' rule \citep{cheriha_descartes_2019} and the intermediate value theorem \citep{hoeven_intermediate_2006}, we are guaranteed a unique real and positive $a_{\rm res}$ value that satisfies Equation~\eqref{eqn:quintic} for a given orbital configuration. 

\section{Results} \label{sec:results}

\subsection{Evection Resonance Configurations during Migration}

\begin{figure*}[t!]
    \centering
    \includegraphics[width=1.02\textwidth]{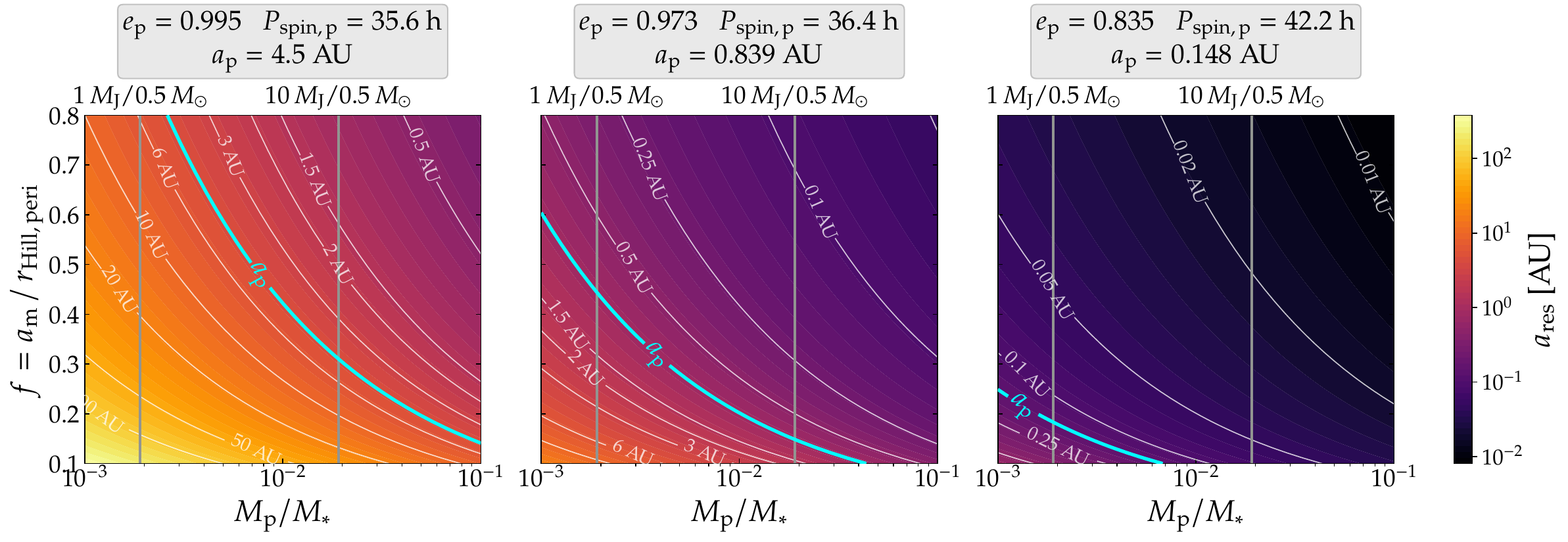}
    \caption{Candidate evection-resonance configurations at three successive stages of the WD--companion tidal migration calculated using Equation~\eqref{eqn:quintic}. Colors and white contours show the companion semi-major axis ($a_{\rm res}$), in AU, satisfying the adopted resonance criterion as a function of mass ratio ($M_{\rm p}/M_{\star}$) and satellite Hill fraction ($f$) at pericenter. Each panel holds the boxed planetary eccentricity $e_{\rm p}$ and spin period $P_{\rm p, \, spin}$ fixed. The cyan curve marks $a_{\rm res} = a_{\rm p}$ where the resonance criterion is satisfied at the displayed orbital configuration. In calculating the contours, the Hill radius is evaluated at $a_{\rm p} = a_{\rm res}$, therefore $f$ on the cyan curve corresponds to the instantaneous lunar Hill fraction. As the companion migrates inward, the resonance locus shifts toward smaller satellite Hill fractions, allowing moons initially interior to the resonance to encounter it even without substantial tidal migration of their own. Gray vertical lines indicate companion masses of $1 \; M_{\rm J}$ and 10 $M_{\rm J}$. All three panels adopt system parameters (e.g., $R_i$, $Q_i$) defined in Section~\ref{subsec:tidal_model}. }
    \label{fig:evection_map} 
\end{figure*}

We evaluate the resonance criterion described in Section~\ref{subsec:evection_criterion} over a range of companion masses and moon orbital separations. Figure~\ref{fig:evection_map} shows three representative stages of the WD--companion tidal evolution, with the corresponding orbital elements and companion spin period indicated above each panel. We construct this reference by solving Equations \eqref{eqn:tidal_migration_a}--\eqref{eqn:tidal_migration_omega} as the companion circularizes, neglecting the moon's feedback on the outer orbit and companion spin. 

Within each panel, the planetary eccentricity ($e_{\rm p}$) and spin period ($P_{\rm p, \, spin} = 2\pi/\Omega_{\rm p}$) are held fixed, while the colors and white contours give the companion semi-major axis ($a_{\rm res}$) at which the resonance criterion is satisfied. The cyan curve highlights $a_{\rm res} = a_{\rm p}$, identifying configurations that satisfy the criterion at the indicated stage of migration. The WD--planet semi-major axis determines which contour is highlighted, while the eccentricity and companion spin determine the underlying contour map. 

As the companion migrates inward, the cyan resonance curve shifts towards smaller Hill fractions at a fixed companion-to-WD mass ratio. It therefore identifies a resonance boundary that can sweep through initially interior satellite orbits as the companion migrates. Capture into the evection resonance is characterized by libration of the resonant angle, 
\begin{equation}
    \phi = 2(\lambda_{\rm p} - \varpi_{\rm m}),
\end{equation}
where $\lambda_{\rm p}$ is the companion's mean longitude, with sustained capture driving the moon's eccentricity growth \citep{spalding_resonant_2016}. However, capture is not required for eccentricity excitation. \citet{xu_disruption_2016} showed that passage through a resonance can also increase eccentricity, with the magnitude of excitation depending on the rate of passage. 

Thus, we regard configurations below the cyan curve as vulnerable to resonance-driven collisions. Whether an individual moon encounters this boundary and reaches a destructive eccentricity depends on its orbital evolution relative to the resonance. Moons initially above the cyan curve are likewise not assured survival, since tidal migration may carry them toward the resonance or lead to a direct collision with their host. We next examine the direction and timescale of satellite migration and compare it with the evolution of WD--companion orbit.  

\subsection{Migration of Satellites}

\begin{figure}[hptb]
    \centering
    \includegraphics[width=\linewidth]{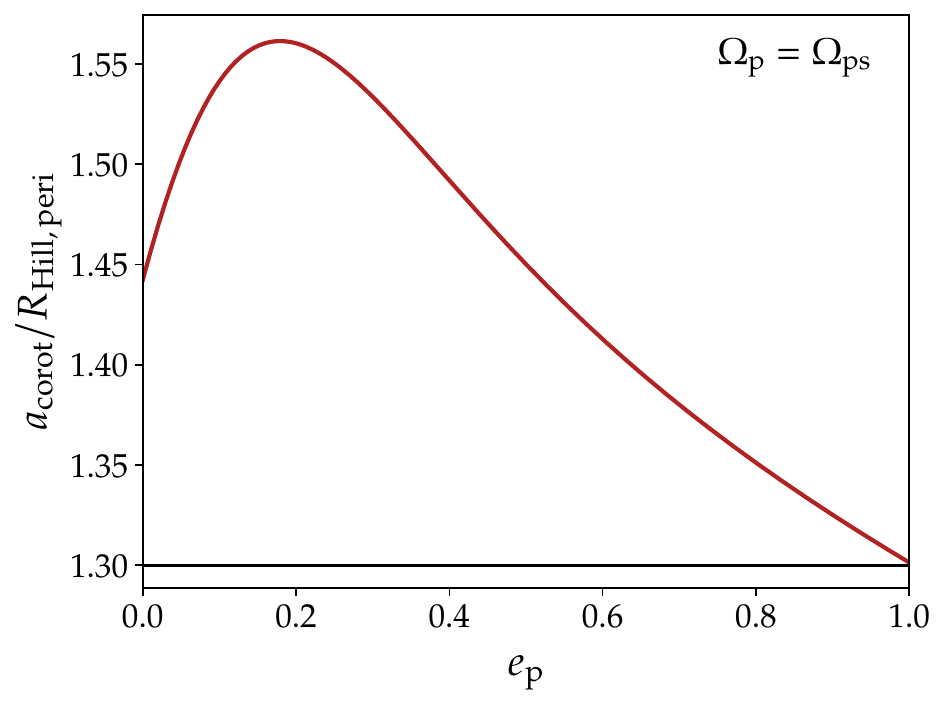}
    \caption{Corotation radius of a planet rotating pseudosynchronously with its host star orbit divided by the Hill radius evaluated at pericenter, as a function of planetary eccentricity. The curve assumes $M_{\rm m} \ll M_{\rm p} \ll M_\star$ and indicates that bound moons will always migrate inwards after planetary spin relaxation. }
    \label{fig:a_corot_r_hill}
\end{figure}

\begin{figure*}[bpth!]
    \centering
    \includegraphics[width=1.02\textwidth]{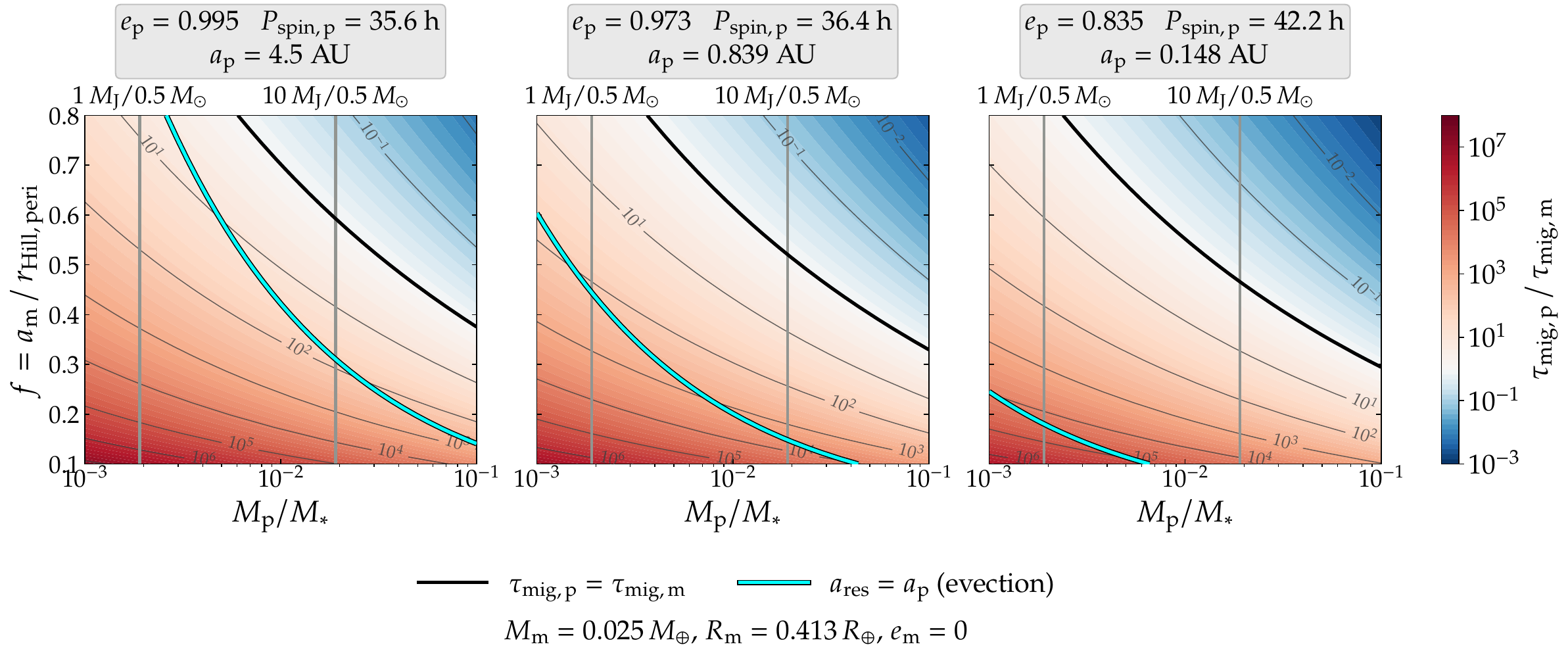}
    \caption{We show the ratio of instantaneous migration timescales, $\mathcal{R} \equiv \tau_{\rm mig, \,p}/\tau_{\rm mig,\, m}$, at the three WD--companion configurations as shown in Figure~\ref{fig:evection_map}. Colors and marked contour lines show $\mathcal{R}$ as a function of companion-to-WD mass ratio and instantaneous Hill fraction at pericenter, with values above unity indicating faster fractional decay of the moon's orbit. The solid black contour marks the boundary where the satellite's migration timescale equals the planet's migration timescale whereas the cyan curve marks the evection resonance condition. Gray vertical lines indicate companion masses of $1 \; M_{\rm J}$ and 10 $M_{\rm J}$. All three panels adopt system parameters (e.g., $R_i$, $Q_i$) defined in Section~\ref{subsec:tidal_model}.}
    \label{fig:pair_timescales}
\end{figure*}

The direction of tidal migration for the moon is determined by the moon's orbital frequency relative to the companion's spin. For a circular, prograde moon, tides raised on the companion drive inward migration when $n_{\rm m} > \Omega_{\rm p}$ and outward migration when $n_{\rm m} < \Omega_{\rm p} $. The corotation radius, where these frequencies are equal, separates these two regimes \citep{barnes_stability_2002}. For a companion rotating pseudosynchronously with its WD orbit (which typically raises the dominant tide on the planet near pericenter), the corotation radius is,
\begin{equation} \label{eqn:a_corot_init}
    a_{\rm corot} = \left( \frac{GM_{\rm p}}{\Omega^2_{\rm ps}} \right)^{1/3}.
\end{equation}
We define the eccentricity-dependent function of the pseudosynchronous rotation rate as, 
\begin{equation} \label{eqn:Ce}
    C(e) \equiv \frac{N(e_{\rm p})}{\omega(e_{\rm p})}.
\end{equation}
In the limit where $M_{\rm m} \ll M_{\rm p} \ll M_\star$, substituting the outer mean motion into Equation (\ref{eqn:a_corot_init}) gives
\begin{equation}
    a_{\rm corot} = a_{\rm p} \left(\frac{M_{\rm p}}{M_\star} \right)^{1/3} C(e_{\rm p})^{-2/3}.
\end{equation}
Expressing this quantity in terms of the Hill radius at pericenter, we have a dimensionless quantity,
\begin{equation} \label{eqn:acorot_rhill}
    \frac{a_{\rm corot}}{R_{\rm Hill, \, peri}} = \frac{3^{1/3}}{C(e_{\rm p})^{2/3} (1-e_{\rm p})}.
\end{equation}
This ratio depends only on the outer orbital eccentricity. Evaluating it over $0 < e_{\rm p} < 1$ gives the curve shown in Figure~\ref{fig:a_corot_r_hill}. The ratio equals $3^{1/3} \simeq 1.44$ for a circular outer orbit and approaches its minimum limiting value of ${\approx}1.30$ as $e_{\rm p} \to 1$. Thus, the corotation radius corresponding to a pseudosynchronously rotating planet always remains outside the Hill radius at pericenter. 

A circular, prograde moon has $a_{\rm m} < R_{\rm Hill, \, peri}$ by definition, and consequently satisfies $a_{\rm m} < a_{\rm corot}$ and $n_{\rm m} > \Omega_{\rm ps}$. Tides raised by the moon on the companion therefore remove angular momentum from the satellite's orbit and drive inward migration. This mechanism is consistent with previous work showing that stellar braking of planetary rotation can drive satellites inward or reverse their initially outward migration \citep{barnes_stability_2002,sasaki_outcomes_2012}. Circular retrograde moons likewise undergo inward tidal migration. Because their orbital motion opposes the companion's rotation, no corotation equilibrium exists at any orbital radius. Then, tides raised on the companion drive orbital decay for as long as the orbit remains retrograde relative to the companion's spin \citep[e.g., Neptune's moon Triton;][]{barnes_stability_2002}. 

\subsection{Pair Migration Timescales} \label{subsec:pair_migration}

Having established the direction of moon migration, we now compare its rate with that of the WD--companion orbit. We define the instantaneous migration timescale for each pair as,
\begin{equation}
    \tau_{{\rm mig}, \, i} = \left|\frac{a_i}{\dot{a}_i} \right|
\end{equation}
and consider their ratio, 
\begin{equation}
    \mathcal{R} \equiv \frac{\tau_{\rm mig, \, p}}{\tau_{\rm mig,\, m}}.
\end{equation}
Values with $\mathcal{R} > 1$ indicate that the satellite orbit contracts more rapidly than the outer orbit, whereas $\mathcal{R} < 1$ identifies configurations with comparatively slow moon migration. We use this comparison to identify configurations potentially favorable to satellite survival during planet migration. A stronger separation, $\mathcal{R} \ll 1$, implies little satellite evolution over the migration timescale of the planet around the WD, potentially providing stability over the timescales relevant for habitability. 

In Figure~\ref{fig:pair_timescales}, we evaluate this ratio at the same three outer orbit configurations used in Figure~\ref{fig:evection_map}, retaining the cyan curve corresponding to the evection resonance for reference. At the initial configuration (Panel 1), the moon's migration is faster for compact satellite orbits whereas sufficiently wide satellites around massive companions have $\mathcal{R} > 1$. At $f=0.5$ for example. the timescale ratio is $\approx 1.5 \times 10^2$ for $1\; M_{\rm J}$ and $\approx 3$ for the $10\; M_{\rm J}$ planet. At this configuration, $\mathcal{R} > 1$ throughout $f \le 0.5$ for companion masses below $\sim 20\; M_{\rm J}$ (which is no longer in the planet regime). Thus, even moons initially exterior to the evection resonance boundary can undergo substantial inward migration relative to the evolution of the outer orbit. 

Increasing either the satellite Hill fraction or the companion mass reduces this timescale ratio. The dependence on satellite separation is particularly strong since
\begin{equation}
    \tau_{\rm mig,\, m} \propto a_{\rm m}^{13/2}
\end{equation}
using our adopted tidal prescription (in the limit that $\Omega_{\rm p} \ll n_{\rm m}$) at fixed masses, radii, and tidal parameters. This dependence is also consistent with the circular satellite migration timescale in Equation (7) of \citet{barnes_stability_2002}. More massive companions also have larger pericenter Hill radii, placing moons at greater physical separations for a given $f$. At our initial configuration, the $\mathcal{R} = 1$ boundary lies at $f \simeq 0.59$ for a $10\;M_{\rm J}$ companion, and at $f \simeq 0.38$ at the right edge of the first panel ($\sim 52 \; M_{\rm J}$) in Figure~\ref{fig:pair_timescales}. At later migration stages, this boundary shifts toward smaller Hill fractions. 

The configurations favored by the tidal comparison must also remain stable against perturbations from the WD at pericenter. Increasing the lunar separation lengthens its tidal migration timescale but places it close to the pericenter Hill radius boundary, where stellar perturbations become more consequential \citep{hamilton_orbital_1992}. More broadly, \citet{trani_ominous_2020} found both satellite inspiral and escape during binary-driven high-eccentricity migration. We therefore next examine whether the outer satellite configurations favored by the timescale comparison can withstand WD pericenter passages. 

\subsection{Pericenter Passages}

\begin{figure*}[bpth!]
    \centering
    \includegraphics[width=1.02\textwidth]{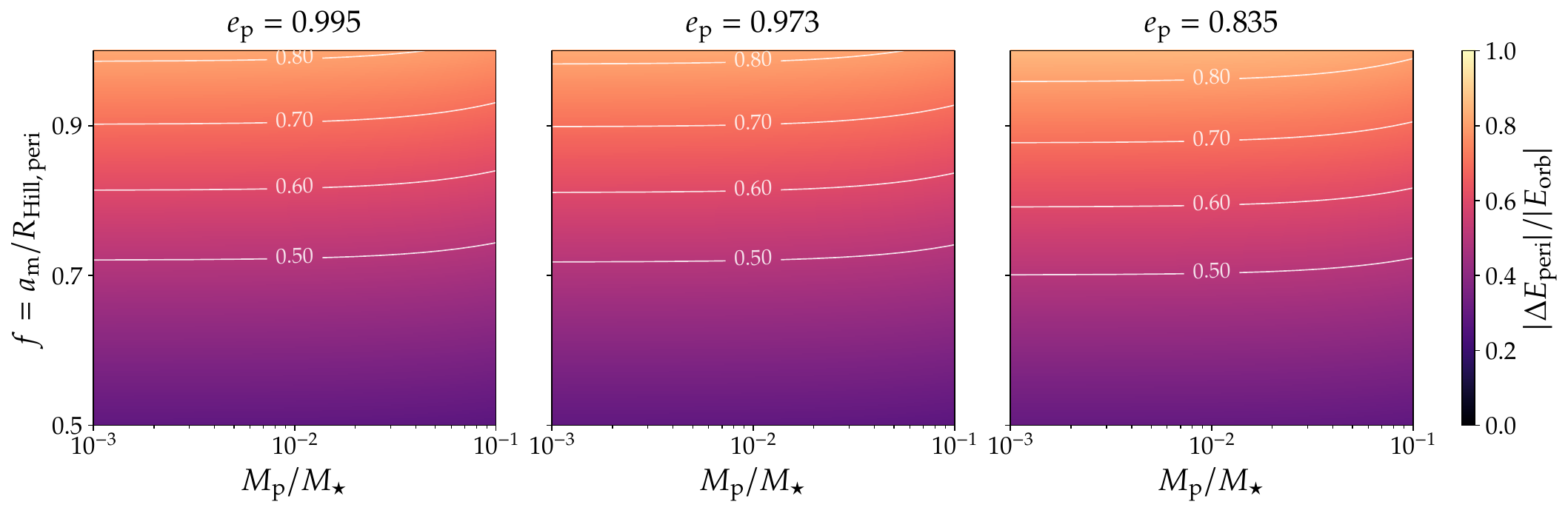}
    \caption{Tidal work exerted by the WD on the satellite orbit during pericenter passage, normalized by the moon--planet orbital binding energy. Colors and white contours show the $|\Delta E_{\rm peri}|/|E_{\rm orb}|$ as a function of planet-to-WD mass ratio, and satellite separation. The panels adopt the same planetary eccentricities as in Figures~\ref{fig:evection_map} and \ref{fig:pair_timescales}. }
    \label{fig:orbital_energy}
\end{figure*}

The differential tidal acceleration across the moon's orbit is \citep{heggie_effect_1996,stone_consequences_2013}, 
\begin{equation}
    \ddot{r}_{\rm tide} \sim \frac{GM_\star}{a_{\rm p}^3(1-e_{\rm p})^3} a_{\rm m},
\end{equation}
where we neglect an orientation-dependent coefficient of order unity. The strongest tidal forcing is at pericenter, with the velocity being entirely tangential, 
\begin{equation}
    v_{\rm p, \, peri} \simeq a_{\rm p} (1-e_{\rm p}) \dot{\nu}_{\rm p, \, peri},
\end{equation}
and consequently,
\begin{equation}
\begin{split}
        t_{\rm peri, \, p} & \sim \frac{r_{\rm p, \, peri}}{v_{\rm p, \, peri}} \simeq 
    \frac{1}{\dot{\nu}_{\rm p, \, peri}} \\ 
    & = \sqrt{\frac{a_{\rm p}^3 (1-e_p)^3}{G(M_\star + M_p)(1+e_{\rm p})}}.
\end{split}
\end{equation}
The moon's orbital speed, neglecting its mass relative to the companion, 
\begin{equation}
    v_{\rm m} = \sqrt{\frac{GM_{\rm p}}{a_{\rm m}}},
\end{equation}
thus the work done on the moon's orbit during pericenter passage is, 
\begin{equation}
    |\Delta E_{\rm peri}| \sim M_{\rm m} \cdot\ddot{r}_{\rm tide} \cdot  v_{\rm m} \cdot    t_{\rm peri}.
\end{equation}
Comparing this with the initial orbital binding energy, 
\begin{equation}
    |E_{\rm orb}| = \frac{GM_{\rm p} M_{\rm m}}{2a_{\rm m}}
\end{equation}
and substituting Equation~\eqref{eqn:a_m} gives
\begin{equation}
    \frac{\Delta E_{\rm peri}}{|E_{\rm orb}|} \sim \frac{2f^{3/2}}{\sqrt{2(1+M_p/M_\star)(1+e_{\rm p})}}.
\end{equation}
We present the resulting ratios as a function of planet-to-star mass ratio and satellite separation in Figure~\ref{fig:orbital_energy}, showing a strong dependence on satellite separation and comparatively weak dependencies on mass ratio and outer eccentricities across the three panels. Across the three displayed configurations, satellites at $f = 0.8$--0.9 have $\Delta E_{\rm peri} \approx 56\%$--73\% of their binding energies. For $f \gtrsim 0.85$, work done at pericenter exceeds 60\% of the orbital binding energy throughout the plotted mass-ratio range in every panel. Thus, the outermost satellite configurations experience gravitational perturbations comparable to their orbital binding energies. 

Close WD passages have also been identified as a mechanism for liberating moons in systems undergoing planetary scattering \citep{payne_liberating_2016}. Our estimate thus shows that wide moons with comparatively long tidal lifetimes are susceptible to substantial orbital changes during pericenter passage, leading to the loss of exomoons either via collision or ejection. 

\section{Discussion} \label{sec:discussion}

Our analysis identifies competing requirements for survival of moons around migrating gas giants orbiting WDs: compact satellite orbits are relatively sheltered from stellar perturbations but undergo rapid inward tidal evolution, whereas wider orbits offer longer tidal lifetimes at the expense of greater exposure to the WD during pericenter passages. Evection provides an additional potential loss channel between these regimes. Together, these constraints make the preservation of primordial satellite systems difficult under our adopted assumptions.

 Several mechanisms can establish the eccentric planetary configurations from which this evolution begins. \citet{munoz_kozai_2020} identified conditions under which vZLK-driven migration can reproduce WD 1856\,b while preserving the planet and completing circularization within the WD cooling age. \citet{oconnor_enhanced_2021} showed that the system's quadruple architecture can broaden the range of initial inclinations permitting extreme eccentricity excitation. More recently, \citet{batygin_terminal_2026} demonstrated that stochastic, asymmetric stellar mass loss \citep{fuller_white_2026} can destabilize planetary architectures and initiate scattering during and after giant-branch evolution. Our calculations pertain to the final survival step: whether a moon that has survived the preceding stellar evolution and excitation of its companion onto a highly eccentric orbit can also survive the subsequent tidal migration. 

 The closest comparison to this work is the study of binary-driven high-eccentricity migration by \citet{trani_ominous_2020}. Their calculations illustrate that moons can actively influence migrations since sufficiently massive satellites induce apsidal precession that can suppress the eccentricity excitation, while systems in which migration proceeds commonly lose their moons. Our approach complements these results by isolating the constraints on an already eccentric companion in the three-body scenario. The resulting competition between tidal decay and stellar perturbations supports the broader conclusion that satellite systems often do not survive planetary migration. 

\citet{gong_effect_2013} and \citet{hong_innocent_2018} found substantial depletion of satellite populations during planet--planet scattering, with \citet{hong_innocent_2018} showing that compact moons (${\sim}0.1\, R_{\rm Hill}$) are preferentially retained. Under our subsequent tidal prescription, however, those compact configurations experience the fastest inward migration. Thus, the moons most protected by the initial dynamical excitation are likely to be destroyed during the later tidal circularization process. 

Earlier studies of satellite tides are directly connected to this idea by the role that planetary rotation plays in causing the inward migration of moons. \citet{barnes_stability_2002} and \citet{sasaki_outcomes_2012} showed how stellar braking and exchange of angular momentum between planetary spin and satellite orbits can produce inward migration or reverse initially outward migration. Our corotation argument identifies a regime in which maintained pseudosynchronous rotation places prograde moons on the inward-migrating side of corotation. This differs from the disk-driven migration pathway considered by \citet{namouni_fate_2010}, where the planetary Hill sphere contracts while satellite separations remain unchanged. Since the pericenter distance increases by about a factor of two as the outer orbit's eccentricity decreases, maintaining a sufficiently long-lived satellite orbit within a restricted pericenter is challenging. 

However, our conclusions depend on the adopted tidal dissipation and planetary spin prescriptions. We use the weak-friction equilibrium-tide model with fixed time lags while our corotation argument assumes maintained pseudosynchronous planetary rotation. Although the CTL equations retain their full eccentricity dependence, this treatment does not capture the excitation and damping of planetary oscillation modes \citep[e.g.,][]{farhat_rattling_2026}. At sufficiently close pericenter passages, repeated excitation of these modes can drive chaotic tidal evolution (via $f$-mode excitation) and rapid early orbital shrinkage \citep{vick_chaotic_2019,veras_tidal_2019}. Such a phase could shorten the time available for satellite inspiral and alter resonance-crossing rates, although subsequent circularization remains subject to the uncertain tidal dissipation. Moreover, our separate examinations of tidal migration, evection, and pericenter perturbations do not fully capture their coupled evolution. Future work should involve long-term three-body integrations that couple orbital and spin evolution and incorporate frequency-dependent, wave-like tides \citep{ogilvie_tidal_2014} when appropriate. Accurately resolving resonance encounters and modeling pericenter passages with more sophisticated methods such as chaotic diffusion (e.g., \citealt{yang_analytical_2025}; E. Yang \& Y. Klein 2026, in preparation) may also be useful. Modeling such effects on more massive companions (such as brown dwarfs) and larger satellite Hill fractions favored by our timescale comparison may help establish satellite survival in a regime where satellite tidal decay is slower but stability against stellar perturbations remains uncertain. 

Satellite depletion could also have consequences beyond the loss of moons from their original companions. \citet{payne_liberating_2016} also found that satellites ejected during scattering can reach smaller WD pericenters than their host planets, potentially contributing atmospheric pollution. \citet{trierweiler_exomoons_2022} subsequently modeled the contribution of moons to observable WD pollution. Migration-driven satellite loss could supply material to such pathways, although determining its contribution requires detailed modeling of the trajectories of such liberated moons. 

The detection of a moon around a close WD companion could therefore provide a diagnostic of its migration history, potentially challenging the high-eccentricity pathways considered here, extending the argument proposed for hot Jupiters by \citet{trani_ominous_2020}. Distinguishing primordial retention from subsequent capture of moons would be essential to interpreting such a discovery. For potentially habitable WD moons, the relevant question consequently extends beyond whether their final orbits permit temperate conditions by considering the evolutionary histories that allow them to reach those environments. 

\begin{acknowledgements}
R.S.N. thanks Alessandro Trani for helpful discussions regarding the use of the \textsc{tsunami} code, and Sarah Millholland and Ritika Sethi for valuable conversations about this work. Support for this
research was provided by the University of Wisconsin--Madison Office of the Vice Chancellor for
Research with funding from the Wisconsin Alumni Research Foundation.
While the original code for the calculations was developed by the authors, Anthropic (Opus 5) was used to refactor the code to run efficiently on a large parameter space and help generate figures. OpenAI (GPT-6 Astra) was used for grammar and for suggestions to enhance clarity in this text. This work was conducted at the University of Wisconsin-Madison, which is located on occupied ancestral land of the Ho-Chunk people, a place their nation has called Teejop since time immemorial. In an 1832 treaty, the Ho-Chunk were forced to cede this territory. The university was founded on and funded through this seized land; this legacy enabled the science presented here. 
\end{acknowledgements}

\bibliography{references}{}
\bibliographystyle{aasjournalv7}

\end{CJK*}
\end{document}